\documentclass{article}
\usepackage{spconfa4,amsmath,graphicx}

\usepackage{bm,amssymb,siunitx}
\usepackage{booktabs}

\usepackage[hidelinks = true]{hyperref}

\title{Residual Learning for Neural Ambisonics Encoders}
\name{%
  \begin{tabular}{@{}c@{}}
  Thomas Deppisch$^{*\dag}$, Yang Gao$^{*}$, Manan Mittal$^{*\ddag}$, Benjamin Stahl$^{*}$,\\
  Christoph Hold$^{*}$, David Alon$^{*}$, Zamir Ben-Hur$^{*}$
  \end{tabular}
}

\address{$^{*}$Reality Labs Research, Meta, WA 98052, USA\\
$^\dag$Chalmers University of Technology, 412 96 Gothenburg, Sweden\\
$^{\ddag}$Stony Brook University, NY 11794, USA}
\begin{document}
\ninept
\maketitle
\begin{abstract}
Emerging wearable devices such as smartglasses and extended reality headsets demand high-quality spatial audio capture from compact, head-worn microphone arrays. Ambisonics provides a device-agnostic spatial audio representation by mapping array signals to spherical harmonic (SH) coefficients. In practice, however, accurate encoding remains challenging. While traditional linear encoders are signal-independent and robust, they amplify low-frequency noise and suffer from high-frequency spatial aliasing. On the other hand, neural network approaches can outperform linear encoders but they often assume idealized microphones and may perform inconsistently in real-world scenarios. To leverage their complementary strengths, we introduce a residual-learning framework that refines a linear encoder with corrections from a neural network. Using measured array transfer functions from smartglasses, we compare a UNet-based encoder from the literature with a new recurrent attention model. Our analysis reveals that both neural encoders only consistently outperform the linear baseline when integrated within the residual learning framework. In the residual configuration, both neural models achieve consistent and significant improvements across all tested metrics for in-domain data and moderate gains for out-of-domain data. Yet, coherence analysis indicates that all neural encoder configurations continue to struggle with directionally accurate high-frequency encoding.
\end{abstract}
\section{Introduction}\label{sec:intro}
Spatial audio technologies have become increasingly popular for immersive applications ranging from virtual and augmented reality to gaming and multimedia entertainment. Among existing spatial audio formats, Ambisonics offers unique advantages through its ability to decouple the recording, processing, and playback stages of the audio pipeline~\cite{Zotter2019}. This decoupling facilitates the application of the same processing algorithms across different recording devices provided that an Ambisonic encoder that delivers a high-quality Ambisonics stream exists for each hardware configuration. Moreover, Ambisonics rendering routines support different playback systems including loudspeaker arrays and personalized headphone playback using head-related transfer functions (HRTFs) without affecting the recording and processing pipelines.

A key challenge to facilitate the Ambisonics processing framework lies in accurately encoding multichannel microphone array signals into the spherical harmonic (SH) domain. Traditional approaches rely on linear least-squares filters~\cite{Moreau2006a,Politis2017a}. Although several studies have improved these methods for head-worn arrays~\cite{Ahrens2022,Bastine2022a,Gayer2025,Gayer2025a}, they suffer from fundamental limitations imposed by the physical constraints of microphone arrays. Compact arrays with small apertures, which are necessary for wearable devices, require large encoder gains to achieve low-frequency directional resolution, increasing the risk of noise amplification and distortion~\cite{Daniel2004}. At high frequencies, spatial aliasing occurs due to limited spatial sampling~\cite{Rafaely2007}.

Signal-dependent approaches attempt to overcome these issues by estimating parameters such as the source direction of arrival and by processing direct and reverberant sound components separately~\cite{Schorkhuber2017a,Politis2018a,Mccormack2022d}. Recent work has also explored deep neural networks (DNNs) for signal-dependent Ambisonic encoding, and several studies have reported improvements over linear methods~\cite{Heikkinen2024,Heikkinen2025a,Qiao2025}. However, existing neural approaches have been evaluated using idealized omnidirectional microphones and image-source simulations of shoebox rooms, limiting their applicability to real-world scenarios. Furthermore, a comprehensive analysis of individual Ambisonics components of neural encoders alongside linear methods has not been performed in the literature, so that a detailed performance comparison is missing.

This paper addresses these limitations by evaluating neural Ambisonic encoder performance across channels and frequencies while considering realistic acoustic behavior of wearable hardware via measured array transfer functions (ATFs) from smartglasses and comparing their behavior with in-domain and out-of-domain data. We compare two neural architectures: a UNet-based encoder from the literature~\cite{Heikkinen2024,Heikkinen2025a} and a new recurrent attention model. The results show that both neural encoders exhibit variations in performance across frequency and SH channel, and that they tend to underperform the linear encoder at low frequencies.

Based on these observations, we propose a residual learning framework that combines the strengths of linear and neural methods. The neural network is trained to learn residual corrections that improve the output of the linear encoder rather than replacing it. While it is shown that the residual configuration consistently improves the results independently for both tested network architectures, our analysis reveals that none of the neural network configurations meaningfully resolves the limitations in high-frequency directional accuracy.

\section{Problem Setup}\label{sec:problem}
Using plane-wave decomposition, the sound pressure ${\bm p(t,f) \in \mathbb{C}^{M}}$ at time $t$ and frequency $f$ picked up by an array of $M$ microphones can be represented as a continuum of plane waves with a corresponding directional amplitude density. The same can be expressed through a linear transform using SHs as basis functions, with the SH-domain plane wave density coefficients $a_n^m$ of order $n$ and degree $m$, and the array response vector ${\bm d_n^m \in \mathbb{C}^{M}}$~\cite{Poletti2000},
\begin{equation}
    \bm p(t,f) = \sum_{n=0}^\infty \sum_{m=-n}^n \bm d_n^m(f) \,a_n^m(t,f) \, .
\end{equation}
For spherical arrangements of ideal microphones, the array vector $\bm d_n^m$ can be analytically described by the product of the SHs ${\bm y_n^m \in \mathbb{R}^{M}}$ evaluated at the microphone directions and the radial terms $b_n^m$, so that ${\tilde{\bm d}_n^m(f) = \bm y_n^m \, b_n^m(f)}$~\cite{Rafaely2019}. For arbitrary arrays, $\bm d_n^m$ can be estimated in a least-squares sense from measured or simulated ATFs for a grid of directions~\cite{Politis2017a}.
Note how the SH-domain signal model decomposes the sound pressure into contributions due to the sound field $a_n^m$ and the microphone array $\bm d_n^m$, making Ambisonics a preferred intermediate format for device-agnostic processing, storage and rendering of spatial audio.

Ambisonics encoders aim to recover a finite subset ${\hat{\bm a} \in \mathbb{C}^{(N+1)^2}}$ of the sound field coefficients $a_n^m$ for orders $n \in [0,N]$ and degrees $m\in [-n,n]$ from the captured sound pressure $\bm p$, effectively removing the influence of spatial sampling with the microphone array. They achieve this with the encoder matrix $\bm E \in \mathbb{C}^{(N+1)^2 \times M}$ so that
\begin{equation}\label{eq:p-encoding}
    \hat{\bm a}(t,f) = \bm E(t,f)\, \bm p(t,f) \, .
\end{equation}
Conventional linear encoders compute $\mathbf E$ as the least-squares optimal solution to~\eqref{eq:p-encoding} by substituting $\bm p$ with measured anechoic ATFs for a grid of incidence directions and $\mathbf a$ with the SHs evaluated at those directions~\cite{Moreau2006a,Politis2017a}.
In contrast, neural network-based methods learn a non-linear mapping from $\bm p$ to $\bm a$ using supervised training where both quantities are known.

Once encoded, Ambisonics signals can be flexibly rendered for loudspeaker arrays or converted to binaural headphone signals~\cite{Zotter2019}.

\section{DNN Methods}\label{sec:methods}
\subsection{DNN Architectures}
Previous studies on neural Ambisonics encoding have explored two types of architectures~\cite{Heikkinen2024,Heikkinen2025a,Qiao2025}. The first group employs UNet-based convolutional networks with skip connections~\cite{Ronneberger2015}, while the second follows a two-stage design that first estimates virtual loudspeaker signals and then derives Ambisonics coefficients. Both approaches operate in the short-time Fourier transform (STFT) domain and treat the real and imaginary parts of the input as separate channels.

\begin{figure}
    \centering
    \includegraphics[trim=0cm 0cm 0cm 0cm,clip,width=0.6\linewidth]{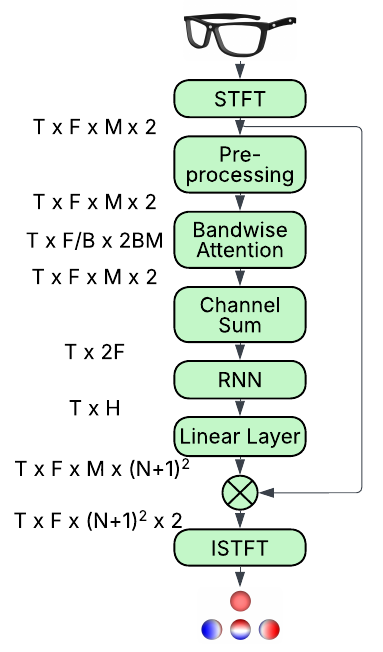}
    \caption{Proposed neural Ambisonics encoder architecture.}
    \label{fig:ambinet}
\end{figure}

In the present contribution, we aim at providing more general insights into neural Ambisonics encoding by comparing two contrasting model architectures with and without a residual learning configuration to the conventional linear encoder. The first model from~\cite{Heikkinen2024} includes a frequency-specific pre-processing stage using a 2D convolution per frequency bin before processing with a UNet. We propose a second architecture designed to contrast the design choice of the first one by combining an attention mechanism and a recurrent neural network (RNN) as illustrated in Fig.~\ref{fig:ambinet}. 

The proposed architecture first uses a frequency-specific pre-processing with a $3 \times 3$ convolution kernel per frequency as in~\cite{Heikkinen2024} which is then being followed by an attention block with a single attention head. The attention mechanism operates on reshaped feature vectors whose embedding dimension spans $B=5$ adjacent frequency bins combined with all input channels and real/imaginary components, enabling the network to learn frequency-dependent relationships across the microphone array.

After attention processing, the multichannel features are combined through summation, reducing the representation to frequency-time features.
These are passed to a two-layer (unidirectional) gated recurrent unit (GRU) with a hidden dimension of $H=512$, processing the temporally-sequenced features to capture temporal dependencies.
A final linear layer maps the GRU hidden states to frequency-specific encoding weights for each input-output channel pair. These real-valued weights are applied as encoding matrix to the original STFT coefficients to produce the Ambisonics signals. 

Note that the UNet-based architecture from~\cite{Heikkinen2024} uses a complex-valued encoding matrix. Extending the proposed model to use complex weights yielded only marginal improvements while substantially increasing computational cost. Therefore, only the real-valued version is considered here. Both architectures use a comparable number of parameters, with 8.1~million for the UNet-based model and 7.9~million for the proposed network.

\subsection{Loss Functions}
The UNet-based model from~\cite{Heikkinen2024} uses a composite loss function with frequency-dependent weights per loss component. For low frequencies, it uses a frequency-domain mean absolute error (MAE), for low and mid frequencies a coherence loss, and at high frequencies an energy preservation component inspired by diffuse-field equalization typically done for linear encoders.

Similary, we propose a loss function composed of the STFT-domain mean absolute error (MAE),
\begin{equation}
    \text{MAE}(f) = \frac{1}{T(N+1)^2} \sum_{t=1}^T \sum_{n=0}^N \sum_{m=-n}^n \left| a_{n}^m(t,f) - \hat{a}_{n}^m(t,f) \right| \, ,
\end{equation}
and the SH coherence loss 
\begin{equation}\label{eq:coherence}
    \mathcal{C}_n^m(f) = 1- \frac{\left| \sum_{t=1}^T a_{n}^m(t,f)^*\, \hat{a}_{n}^m(t,f) \right|^2}{ \sum_{t'=1}^T \left| a_{n}^m(t',f)\right|^2 \, \sum_{t''=1}^T \left| \hat{a}_{n}^m(t'',f)\right|^2}\, ,
\end{equation}
to promote an accurate reproduction of SH directivity patterns. The total loss is defined as
\begin{equation}\label{eq:loss}
    L = \frac{1}{F} \sum_{f=1}^F \left( \alpha\, \text{MAE}(f) + \frac{\beta(f)}{N+1} \sum_{n=0}^N \sum_{m=-n}^n \frac{1}{2n+1} \, \mathcal{C}_n^m(f) \right) \, ,
\end{equation}
where $\alpha = 10$, and $\beta(f)$ is inversely proportional to frequency, ensuring approximately equal weighting across octave bands similar to human perception. The SH-order weighting $1/(2n+1)$ normalizes the contribution of each order such that all components within the same order have equal overall influence, i.e., the zeroth-order term $\mathcal{C}_0^0$ contributes as much to the total loss as the combined first-order components ${\mathcal{C}_1^{-1}, \mathcal{C}_1^{0}, \mathcal{C}_1^{1}}$.
No explicit transition frequencies are defined, allowing the loss formulation to remain applicable to a wide range of array configurations without requiring array-specific tuning.

\subsection{Residual Learning}
As shown in the subsequent analysis in Sec.~\ref{sec:results}, neither of the tested neural network architectures consistently outperforms the linear encoder across all SH channels and frequencies. To address this limitation, we propose to adopt a residual learning strategy to leverage strengths of the linear encoder while improving performance using the neural approaches. 

\begin{figure}
    \centering
    \includegraphics[trim=0cm 0.2cm 0cm 0.2cm,clip,width=0.7\linewidth]{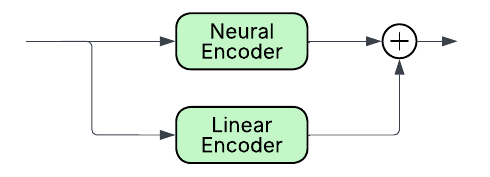}
    \caption{The proposed residual learning framework trains a neural encoder end-to-end to minimize a loss computed from the sum of the neural and linear encoder outputs.}
    \label{fig:residual-enc}
\end{figure}

The proposed approach is illustrated in Fig.~\ref{fig:residual-enc}. The input signal is processed in parallel by the linear and neural encoders, and their time-domain outputs are summed to produce the final Ambisonics signal while ensuring that the processing delays of the linear and neural encoders match. The neural network is trained end-to-end to minimize a loss function computed from the combined output. We also tested learning optimal summation weights for each SH component but found no improvement over simple summation.

\section{Evaluation}\label{sec:eval}
\subsection{Datasets}
A dataset of 80,000 scenes was simulated by generating room impulse responses (RIRs) for one to five randomly positioned sources in 80,000 shoebox-shaped rooms using pyroomacoustics~\cite{Scheibler2018}. The RIRs were computed using the image source method up to order 40, with randomly generated dimensions between $3 \times 3 \times 2$~\si{m} and $10 \times 10 \times 5$~\si{m} and absorption coefficients between 0.2 and 0.95 for all surfaces, at a sampling rate of \SI{48}{\kilo\hertz}.
To model the acoustic response of a wearable microphone array, the simulation included measured ATFs of a five-microphone smartglasses array (one microphone on the nose bridge and two on each temple) for a dense grid of source directions. An ideal SH receiver was also simulated to provide ground-truth data for training and evaluation.

Array signals of \SI{8}{\second} duration were generated by convolving each RIR with randomly selected audio samples from different datasets. For training and validation, audio was drawn from the DNS dataset~\cite{Dubey2022}
and the FUSS dataset~\cite{Wisdom2021a},
with 60\% speech (from DNS) and 40\% non-speech signals (from DNS and FUSS). For testing, the audio consisted of 60\% speech from the EARS dataset~\cite{Richter2024}
and 40\% non-speech from the FUSS dataset. The resulting dataset comprises \SI{178}{\hour} of audio in 80,000 reverberant scenes and was split into 80\%/10\%/10\% for training, validation, and testing.

To evaluate out-of-domain generalization, a second test dataset of 20~rooms was generated using the Treble hybrid simulator\footnote{https://www.treble.tech/}, 
which combines wave-based simulation below \SI{3.2}{\kilo\hertz} with ray tracing above and includes frequency-dependent surface absorption and furniture as scattering objects. The simulated rooms represent typical real-world environments such as meeting rooms, living rooms, bedrooms, restaurants, and full apartments. For each room, 15 receiver positions were simulated, each with one to five source positions. RIRs were convolved with audio signals in the same manner as the in-domain test set. This dataset contains \SI{40}{\minute} of audio across 300 scenes and provides realistic scenarios to test method generalization.

\subsection{Model Setup and Training}
All encoders used a maximum SH order of $N=1$ which is the maximum feasible order for linear encoding with $M=5$ microphones. The linear encoder used diffuse field equalization as in~\cite{Mccormack2022d} and a \SI{20}{dB} gain limitation to limit low-frequency noise amplification. 

All neural models were trained for 250~epochs on \SI{1}{\second} input segments with a batch size of 128. The STFT processing used a frame size of 768~samples with 50\% overlap. The checkpoint achieving the lowest validation loss was used for evaluation with the test sets. Input signals were normalized per frequency according to the variance of the training set.

We evaluate six neural configurations against the linear encoder:
\begin{itemize}
    \item UNet~\cite{Heikkinen2024} with the original loss and its residual variant (UResNet)
    \item UNet (mod.) trained with the loss from~\eqref{eq:loss} and its residual variant (UResNet~(mod.))
    \item The proposed attention-RNN architecture in Fig.~\ref{fig:ambinet} (AmbiNet) and its residual variant (AmbiResNet)
\end{itemize}
All models were trained using the Adam optimizer with an initial learning rate of $10^{-3}$, which was reduced by a factor of 0.3 whenever the validation loss did not improve for 10 consecutive epochs.

\subsection{Evaluation Metrics}
\begin{table*}[t]
    \footnotesize
    \centering
    \label{tab:results}
    \begin{tabular}{l rrrr c rrrr}
        \toprule
        & \multicolumn{4}{c}{\textbf{In-Domain}} & & \multicolumn{4}{c}{\textbf{Out-of-Domain}} \\
        \cmidrule(lr){2-5} \cmidrule(lr){7-10}
        \textbf{Model} & \textbf{Coh.}$\uparrow$ & \textbf{Mag. Err. (dB)}$\downarrow$ & \textbf{SI-SDR (dB)}$\uparrow$ & \textbf{SPME (dB)}$\downarrow$ & & \textbf{Coh.}$\uparrow$ & \textbf{Mag. Err. (dB)}$\downarrow$ & \textbf{SI-SDR (dB)}$\uparrow$ & \textbf{SPME (dB)}$\downarrow$ \\
        \cmidrule(lr){1-5} \cmidrule(lr){7-10}
        Linear Enc. & $0.44$ & $11.0$ & $-7.4$ & $-0.4$ & & $0.39$ & $13.3$ & $-9.7$ & $9.5$ \\
        \cmidrule(lr){1-5} \cmidrule(lr){7-10}
        UNet \cite{Heikkinen2024} & $0.49$ & $11.0$ & $-6.8$ & $8.4$ & & $0.37$ & $17.9$ & $-11.4$ & $\mathbf{9.8}$ \\
        UResNet & $\mathbf{0.54}^*$ & $\mathbf{8.5}^*$ & $\mathbf{-4.0}$ & $\mathbf{-3.2}$ & & $\mathbf{0.40}^*$ & $\mathbf{13.1}^*$ & $\mathbf{-7.6}$ & $10.4$ \\
        \cmidrule(lr){1-5} \cmidrule(lr){7-10}
        UNet (mod.) & $0.50$ & $9.4$ & $-0.2$ & $7.2$ & & $0.32$ & $15.3$ & $-5.4$ & $8.7$ \\
        UResNet (mod.) & $\mathbf{0.53}$ & $\mathbf{8.7}$ & $\mathbf{1.0}^*$ & $\mathbf{-2.4}$ & & $\mathbf{0.40}^*$ & $\mathbf{14.2}$ & $\mathbf{-4.0}^*$ & $\mathbf{6.5}^*$ \\
        \cmidrule(lr){1-5} \cmidrule(lr){7-10}
        AmbiNet & $0.49$ & $10.4$ & $0.0$ & $2.2$ & & $0.36$ & $\mathbf{13.6}$ & $-5.0$ & $\mathbf{6.5}^*$ \\
        AmbiResNet & $\mathbf{0.52}$ & $\mathbf{8.6}$ & $\mathbf{0.2}$ & $\mathbf{-4.4}^*$ & & $\mathbf{0.40}^*$ & $13.8$ & $\mathbf{-4.9}$ & $8.2$ \\
        \bottomrule
    \end{tabular}
    \caption{Performance of the linear encoder and the three neural encoders, with and without residual learning. Best results per architecture are shown in bold, and overall best results are marked with an asterisk.}
\end{table*}

To assess encoder performance across different aspects, we employ four complementary metrics:
\begin{itemize}
    \item The coherence metric from~\eqref{eq:coherence} quantifies the frequency-dependent ability to accurately reproduce SH directivity patterns and was averaged over channels and frequency-weighted as in~\eqref{eq:loss}.
    \item The mean magnitude spectral error,
\begin{equation}
    S(f) = \frac{1}{T (N+1)^2} \sum_{t=1}^T \sum_{n=0}^N \sum_{m=-n}^n \left | 20 \log_{10} \left( \frac{|a_{n}^m(t,f)|}{|\hat{a}_{n}^m(t,f)|}\right) \right | \, ,
\end{equation}
quantifies frequency-domain magnitude deviations in decibels.
    \item The scale-invariant signal-to-distortion ratio (SI-SDR) measures the time-domain signal-to-distortion ratio while being invariant to scale differences between the reference and estimated signals~\cite{Luo2019a,Roux2019}.
    \item The spatial power map error (SPME)~\cite{Qiao2025},
\begin{equation}
    D(f) = 20 \log_{10} \,\frac{1}{Q} \sum_{q=1}^Q \left| \hat{\Gamma}(\Omega_q) - \Gamma(\Omega_q) \right| \, ,
\end{equation}
characterizes the directional energy distribution across $Q=1296$ uniformly distributed directions on the sphere~\cite{Graf2011}. Here,
\begin{equation}
    \Gamma(\Omega_q) = \sqrt{\sum_{t=1}^T \left( \sum_{n=0}^N \sum_{m=-n}^n Y_n^m(\Omega_q)\, a_{n}^m(t)  \right)^2} \, 
\end{equation}
represents the energy at direction $\Omega_q$ using the real-valued SHs $Y_n^m(\Omega_q)$. This can be interpreted as decoding the Ambisonics signal to a dense spherical loudspeaker array and comparing each loudspeaker's energy to the ground truth.
\end{itemize}

\subsection{Results}\label{sec:results}
\begin{figure}[!t]
    \centering
    \includegraphics[trim=0cm 1.3cm 0cm 0cm,clip,width=\linewidth]{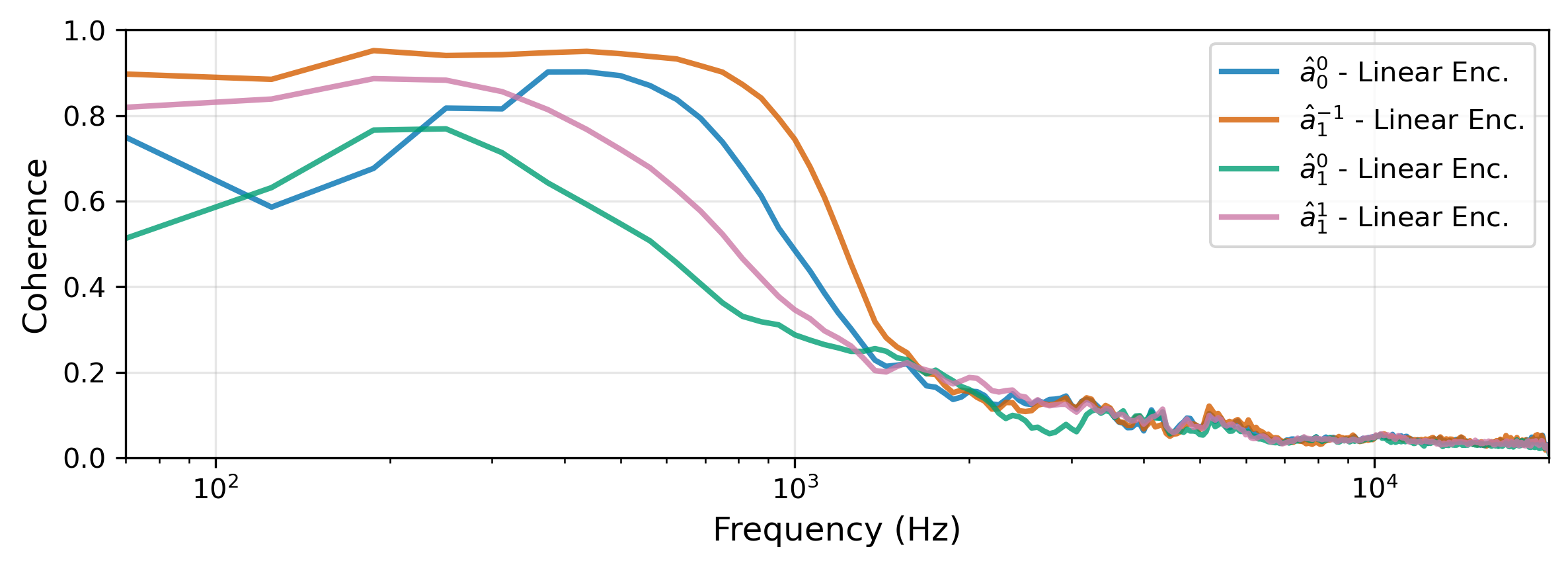}\\
    \includegraphics[trim=0cm 1.3cm 0cm 0cm,clip,width=\linewidth]{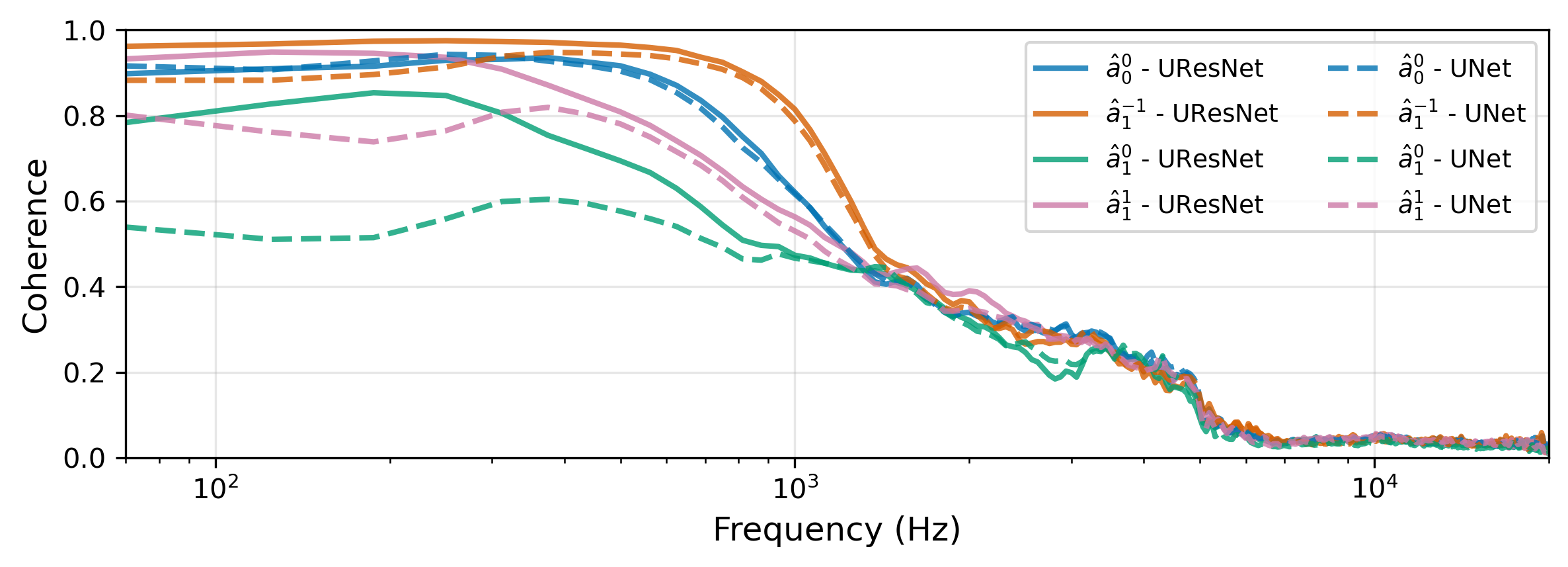}\\
    \includegraphics[trim=0cm 1.3cm 0cm 0cm,clip,width=\linewidth]{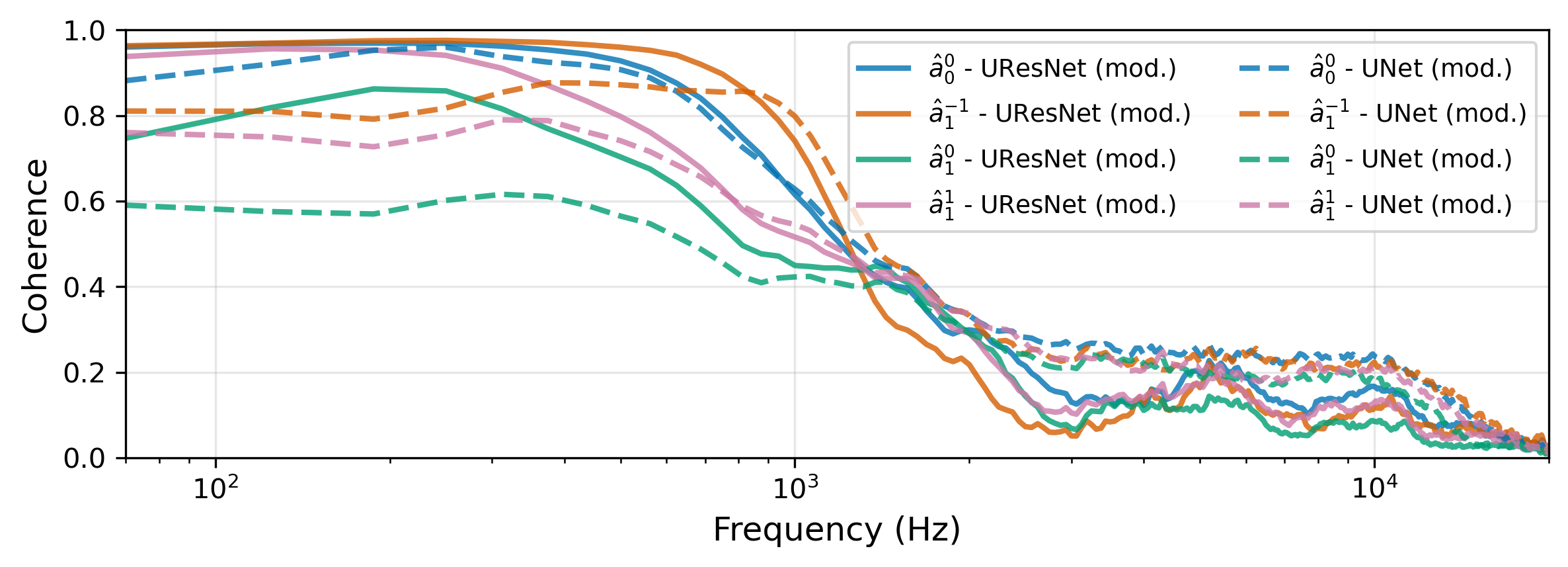}\\
    \includegraphics[width=\linewidth]{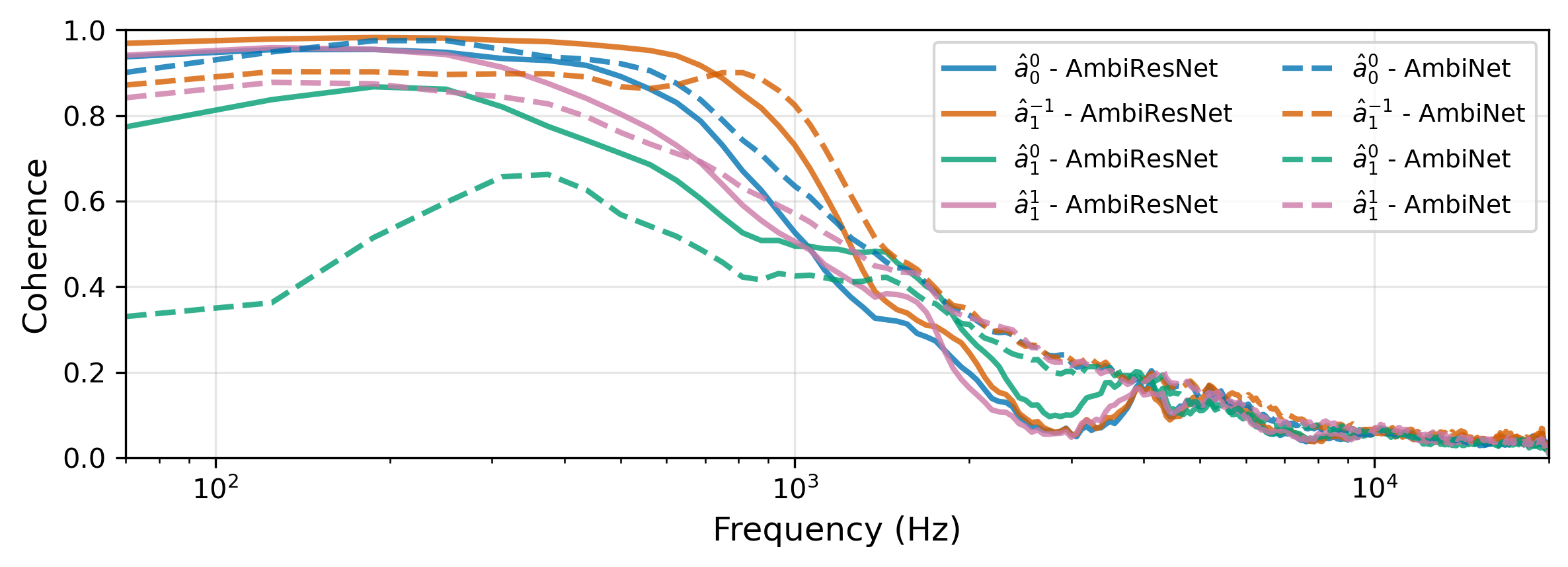}
    \caption{Frequency-dependent coherence for the linear encoder (top) and the neural encoders with residual learning (solid lines) and without residual learning (dashed lines). Higher coherence suggests better reproduction of SH directivity patterns.}
    \label{fig:coh_results}
\end{figure}

Fig.~\ref{fig:coh_results} shows the frequency-dependent coherence per SH channel for the linear encoder and the neural encoders with and without residual learning on the in-domain dataset. The coherence metric characterizes the encoders' ability to capture the directionality of the sound scene and is the principal metric for directional encoding performance. Despite their different architectures, the standalone neural encoders (dashed lines) perform similarly, exceeding the linear baseline (top plot) only in specific frequency bands and channels, for example $\hat{a}_0^0$ below \SI{400}{\hertz}, while the linear encoder remains superior in others, such as $\hat{a}_1^0$ between \SI{100}{\hertz} and \SI{500}{\hertz}. In contrast, the residual variants (solid lines) consistently match or surpass the linear encoder's coherence across all channels and frequencies.

For all encoders, performance varies similarly across SH components and frequencies. For components where the linear encoder exhibits lower performance due to the geometry of the microphone array with microphones around the front and sides of the head, such as the up/down component $\hat{a}_1^0$ and the front/back component $\hat{a}_1^1$, the neural encoders also achieve lower coherence. Additionally, coherence drops toward high frequencies across all configurations, where the linear encoder exhibits spatial aliasing. 

Tab.~\ref{tab:results} summarizes results for all metrics on both datasets. On the in-domain data, residual learning produces consistent improvements over the linear baseline and the corresponding non-residual networks. Coherence increases from $0.44$ (linear) to $0.52$–$0.54$, with UResNet achieving the highest value. The mean magnitude error decreases from $11.0$~dB to $8.5$–$8.7$~dB, again with UResNet best. SI-SDR improves substantially relative to the linear encoder, with UResNet~(mod.) reaching $1.0$~dB compared with $-7.4$~dB for the linear model. SPME is reduced below the linear baseline for all residual models, with AmbiResNet achieving the lowest value at $-4.4$~dB.

Out-of-domain results show smaller but mostly positive gains from residual learning. Non-residual neural models often underperform the linear encoder, whereas residual variants recover and frequently exceed its performance. Coherence increases slightly from $0.39$ (linear) to $0.40$ for all residual models, magnitude error improves modestly to $13.1$~dB for UResNet, SI-SDR improves from $-9.7$~dB to $-4.0$~dB (UResNet (mod.)), and SPME only decreases for UResNet, with UResNet (mod.) and AmbiNet achieving the lowest error of $6.5$~dB.

These results indicate that residual learning improves generalization and reliability relative to standalone neural encoders while retaining the robustness of the linear solution. The coherence drop toward high frequencies, however, suggests that none of the approaches, including the non-linear neural network solutions, can substantially mitigate errors arising from limited spatial sampling. The modest gains on out-of-domain data highlight the benefit of realistic training data, even when residual learning is employed. 
Informal listening indicated that the residual networks sound largely similar to the linear encoder. While minor improvements were observed in some cases, they were insufficient to justify a formal perceptual study, but binaural examples are shared online.\footnote{\url{https://thomasdeppisch.github.io/neural-ambisonics-encoding/}}

\section{Conclusion}
\label{sec:conclusion}
We presented a residual-learning framework for neural Ambisonics encoding that refines the output of a linear least-squares encoder with learned corrections. Our results for a compact microphone array on smartglasses demonstrate that standalone neural encoders do not consistently outperform the linear baseline across SH channels and frequencies, whereas the residual learning strategy provides consistent improvements on both in-domain and out-of-domain datasets, highlighting its enhanced reliability and generalization compared with purely neural approaches. 

Overall, however, the results show that none of the approaches can significantly overcome limitations from sparse spatial sampling with the microphone array. Future work should thus explore network designs for directional super-resolution with a limited number of microphones and may draw inspiration from parametric encoding approaches that aim to identify individual sound sources. The residual learning framework introduced here may prove beneficial for such approaches as well.


\bibliographystyle{IEEEbib}
\bibliography{references}

@inproceedings{Moreau2006a,
    title = {{3D Sound Field Recording with Higher Order Ambisonics – Objective Measurements and Validation of Spherical Microphone}},
    year = {2006},
    booktitle = {Proc. 120th Conv. Audio Eng. Soc.},
    author = {Moreau, Sébastien and Daniel, Jérôme and Bertet, Stéphanie},
    pages = {1--24}
}

@article{Poletti2000,
    title = {{A Unified Theory of Horizontal Holographic Sound Systems}},
    year = {2000},
    journal = {J. Audio Eng. Soc.},
    author = {Poletti, Mark},
    number = {12},
    pages = {1155–1182},
    volume = {48}
}

@inproceedings{Bastine2022a,
    title = {{Ambisonics Capture using Microphones on Head-worn Device of Arbitrary Geometry}},
    year = {2022},
    booktitle = {European Signal Processing Conference},
    author = {Bastine, Amy and Birnie, Lachlan and Abhayapala, Thushara D. and Samarasinghe, Prasanga and Tourbabin, Vladimir},
    pages = {309--313},
    isbn = {9789082797091},
    doi = {10.23919/eusipco55093.2022.9909803},
    issn = {22195491}
}

@inproceedings{Gayer2025a,
    title = {{Ambisonics Encoder for Wearable Array with Improved Binaural Reproduction}},
    year = {2025},
    booktitle = {Forum Acusticum Euronoise},
    author = {Gayer, Yhonatan and Tourbabin, Vladimir and Ben-Hur, Zamir and Alon, David and Rafaely, Boaz},
    url = {http://arxiv.org/abs/2507.04108},
    arxivId = {2507.04108}
}

@book{Zotter2019,
    title = {{Ambisonics, A Practical 3D Audio Theory for Recording, Studio Production, Sound Reinforcement, and Virtual Reality}},
    year = {2019},
    booktitle = {Springer Topics in Signal Processing},
    author = {Zotter, Franz and Frank, Matthias},
    publisher = {Springer},
    doi = {10.1007/978-3-030-17207-7}
}

@article{Gayer2025,
    title = {{Array-Aware Ambisonics and HRTF Encoding for Binaural Reproduction With Wearable Arrays}},
    year = {2025},
    journal = {arXiv:2507.11091},
    author = {Gayer, Yhonatan and Tourbabin, Vladimir and Hur, Zamir Ben and Alon, David Lou and Rafaely, Boaz},
    pages = {1--11},
    url = {http://arxiv.org/abs/2507.11091},
    doi = {arXiv:2507.11091},
    arxivId = {2507.11091}
}

@inproceedings{Politis2017a,
    title = {{Comparing modeled and measurement-based spherical harmonic encoding filters for spherical microphone arrays}},
    year = {2017},
    booktitle = {IEEE Workshop on Applications of Signal Processing to Audio and Acoustics},
    author = {Politis, Archontis and Gamper, Hannes},
    pages = {224--228},
    isbn = {9781538616321},
    doi = {10.1109/WASPAA.2017.8170028}
}

@article{Luo2019a,
    title = {{Conv-TasNet: Surpassing Ideal Time-Frequency Magnitude Masking for Speech Separation}},
    year = {2019},
    journal = {IEEE/ACM Transactions on Audio Speech and Language Processing},
    author = {Luo, Yi and Mesgarani, Nima},
    number = {8},
    pages = {1256--1266},
    volume = {27},
    url = {http://arxiv.org/abs/1809.07454 http://dx.doi.org/10.1109/TASLP.2019.2915167},
    doi = {10.1109/TASLP.2019.2915167},
    issn = {23299304},
    arxivId = {1809.07454}
}

@inproceedings{Richter2024,
    title = {{EARS: An Anechoic Fullband Speech Dataset Benchmarked for Speech Enhancement and Dereverberation}},
    year = {2024},
    booktitle = {Proceedings of INTERSPEECH},
    author = {Richter, Julius and Wu, Yi Chiao and Krenn, Steven and Welker, Simon and Lay, Bunlong and Watanabe, Shinji and Richard, Alexander and Gerkmann, Timo},
    pages = {4873--4877},
    doi = {10.21437/Interspeech.2024-153},
    issn = {19909772},
    arxivId = {2406.06185}
}

@book{Rafaely2019,
    title = {{Fundamentals of Spherical Array Processing}},
    year = {2019},
    author = {Rafaely, Boaz},
    edition = {2nd},
    publisher = {Springer}
}

@inproceedings{Daniel2004,
    title = {{Further Study of Sound Field Coding with Higher Order Ambisonics}},
    year = {2004},
    booktitle = {Proc. 116th Conv. Audio Eng. Soc.},
    author = {Daniel, Jérôme and Moreau, Sébastien},
    pages = {1--14}
}

@inproceedings{Heikkinen2025a,
    title = {{Gen-A: Generalizing Ambisonics Neural Encoding to Unseen Microphone Arrays}},
    year = {2025},
    booktitle = {IEEE Int. Conf. on Acoustics, Speech and Signal Processing (ICASSP)},
    author = {Heikkinen, Mikko and Politis, Archontis and Drossos, Konstantinos and Virtanen, Tuomas},
    pages = {1--5},
    url = {http://arxiv.org/abs/2501.08047},
    isbn = {9798350368741},
    doi = {10.1109/ICASSP49660.2025.10887869},
    issn = {15206149},
    arxivId = {2501.08047}
}

@inproceedings{Dubey2022,
    title = {{ICASSP 2022 Deep Noise Suppression Challenge}},
    year = {2022},
    booktitle = {IEEE Int. Conf. on Acoustics, Speech and Signal Processing (ICASSP)},
    author = {Dubey, Harishchandra and Gopal, Vishak and Cutler, Ross and Aazami, Ashkan and Matusevych, Sergiy and Braun, Sebastian and Eskimez, Sefik Emre and Thakker, Manthan and Yoshioka, Takuya and Gamper, Hannes and Aichner, Robert},
    pages = {9271--9275},
    isbn = {9781665405409},
    doi = {10.1109/ICASSP43922.2022.9747230},
    issn = {15206149}
}

@inproceedings{Qiao2025,
    title = {{Neural Ambisonic Encoding For Multi-Speaker Scenarios Using A Circular Microphone Array}},
    year = {2025},
    booktitle = {IEEE Int. Conf. on Acoustics, Speech and Signal Processing (ICASSP)},
    author = {Qiao, Yue and Kothapally, Vinay and Yu, Meng and Yu, Dong},
    pages = {1--5},
    doi = {10.1109/ICASSP49660.2025.10890048},
    issn = {15206149},
    arxivId = {2409.06954}
}

@inproceedings{Heikkinen2024,
    title = {{Neural Ambisonics Encoding For Compact Irregular Microphone Arrays}},
    year = {2024},
    booktitle = {IEEE Int. Conf. on Acoustics, Speech and Signal Processing (ICASSP)},
    author = {Heikkinen, Mikko and Politis, Archontis and Virtanen, Tuomas},
    pages = {701--705},
    doi = {10.1109/icassp48485.2024.10447425},
    issn = {15206149},
    arxivId = {2401.05916}
}

@article{Graf2011,
    title = {{On the computation of spherical designs by a new optimization approach based on fast spherical Fourier transforms}},
    year = {2011},
    journal = {Numerische Mathematik},
    author = {Gr{\"{a}}f, Manuel and Potts, Daniel},
    number = {4},
    pages = {699--724},
    volume = {119}
}

@article{Mccormack2022d,
    title = {{Parametric Ambisonic Encoding of Arbitrary Microphone Arrays}},
    year = {2022},
    journal = {IEEE/ACM Transactions on Audio, Speech, and Language Processing},
    author = {Mccormack, Leo and Politis, Archontis and Gonzalez, Raimundo and Lokki, Tapio and Pulkki, Ville},
    pages = {2062--2075},
    volume = {30},
    doi = {10.1109/TASLP.2022.3182857}
}

@inproceedings{Politis2018a,
    title = {{Parametric multidirectional decomposition of microphone recordings for broadband high-order ambisonic encoding}},
    year = {2018},
    booktitle = {144th Conv. Audio Eng. Soc.},
    author = {Politis, Archontis and Tervo, Sakari and Lokki, Tapio and Pulkki, Ville},
    pages = {1–10}
}

@inproceedings{Scheibler2018,
    title = {{Pyroomacoustics: A Python package for audio room simulations and array processing algorithms}},
    year = {2018},
    booktitle = {IEEE International Conference on Acoustics, Speech and Signal Processing (ICASSP)},
    author = {Scheibler, Robin and Bezzam, Eric and Dokmanic, Ivan},
    pages = {351–355},
    doi = {10.1109/ICASSP.2018.8461310}
}

@inproceedings{Roux2019,
    title = {{SDR - Half-baked or Well Done?}},
    year = {2019},
    booktitle = {IEEE Int. Conf. on Acoustics, Speech and Signal Processing (ICASSP)},
    author = {Roux, Jonathan Le and Wisdom, Scott and Erdogan, Hakan and Hershey, John R.},
    pages = {626--630},
    isbn = {9781479981311},
    doi = {10.1109/ICASSP.2019.8683855},
    issn = {15206149},
    arxivId = {1811.02508}
}

@inproceedings{Schorkhuber2017a,
    title = {{Signal-Dependent Encoding for First-Order Ambisonic Microphones}},
    year = {2017},
    booktitle = {Proc. of the Annual German Conference on Acoustics (DAGA)},
    author = {Sch{\"{o}}rkhuber, Christian and H{\"{o}}ldrich, Robert},
    pages = {1037--1040}
}

@article{Rafaely2007,
    title = {{Spatial aliasing in spherical microphone arrays}},
    year = {2007},
    journal = {IEEE Transactions on Signal Processing},
    author = {Rafaely, Boaz and Weiss, Barak and Bachmat, Eitan},
    number = {3},
    pages = {1003--1010},
    volume = {55},
    doi = {10.1109/TSP.2006.888896},
    issn = {1053587X}
}

@article{Ahrens2022,
    title = {{Spherical Harmonic Decomposition of a Sound Field Using Microphones on a Circumferential Contour Around a Non-Spherical Baffle}},
    year = {2022},
    journal = {IEEE/ACM Transactions on Audio Speech and Language Processing},
    author = {Ahrens, Jens and Helmholz, Hannes and Alon, David Lou and Amengual Gar{\'{i}}, Sebastià V.},
    pages = {3110--3119},
    volume = {30},
    doi = {10.1109/TASLP.2022.3209940},
    issn = {23299304}
}

@inproceedings{Ronneberger2015,
    title = {{U-Net: Convolutional Networks for Biomedical Image Segmentation}},
    year = {2015},
    booktitle = {Medical Image Computing and Computer-Assisted Intervention – MICCAI},
    author = {Ronneberger, Olaf and Fischer, Philipp and Brox, Thomas},
    pages = {234--241},
    url = {arxiv:1505.04597},
    doi = {arXiv:1505.04597},
    arxivId = {arXiv:1505.04597}
}

@inproceedings{Wisdom2021a,
    title = {{What's all the fuss about free universal sound separation data?}},
    year = {2021},
    booktitle = {IEEE Int. Conf. on Acoustics, Speech and Signal Processing (ICASSP)},
    author = {Wisdom, Scott and Erdogan, Hakan and Ellis, Daniel P.W. and Serizel, Romain and Turpault, Nicolas and Fonseca, Eduardo and Salamon, Justin and Seetharaman, Prem and Hershey, John R.},
    pages = {186--190},
    isbn = {9781728176055},
    doi = {10.1109/ICASSP39728.2021.9414774},
    issn = {15206149},
    arxivId = {2011.00803}
}

\end{document}